\documentclass[aps,prb,preprint,groupedaddress,showpacs,preprintnumbers]{revtex4-1}
\usepackage[usenames]{color}
\usepackage{natbib}
\usepackage{graphicx}
\usepackage{float}
\begin{document}
\definecolor{Red}{rgb}{1.00,0.00,0.00}
\definecolor{Blue}{rgb}{0.24,0.24,0.76}
\newcommand{\AV}{\color{Red}}
\newcommand{\MJC}{\color{Blue}}
\preprint{}


\title{Vacancy in graphene: insight on magnetic properties from theoretical modeling}
\author{A. M. Valencia}
\email[]{valencia@if.usp.br}
\affiliation{Instituto de F\'{\i}sica, Universidade de S\~ao Paulo, CEP 66318, 05315-970, S\~ao Paulo-SP, Brazil}
\author{M. J. Caldas}
\affiliation{Instituto de F\'{\i}sica, Universidade de S\~ao Paulo, CEP 66318, 05315-970, S\~ao Paulo-SP, Brazil}

\date{\today}

\begin{abstract}
	Magnetic properties of a single vacancy in graphene is a relevant and still much discussed problem. The experimental results point to a clearly detectable magnetic defect state at the Fermi energy, while calculations based on density functional theory (DFT) yield widely varying results for the magnetic moment, in the range of $\mu=1.04-2.0$ $\mu_{B}$. We present a multi-tool \textit{ab initio} theoretical study of the same defect, using two simulation protocols for a defect in a crystal (cluster and periodic boundary conditions) and different DFT functionals - bare and hybrid DFT, mixing a fraction of Hartree-Fock exchange (XC). Our main conclusions are two-fold: First, we find that due to the $\pi$-character of the Fermi-energy states of graphene, inclusion of XC is crucial and for a single isolated vacancy we can predict an integer magnetic moment $\mu=2\mu_{B}$. Second, we find that due to the specific symmetry of the graphene lattice, periodic arrays of single vacancies may provide interesting diffuse spin-spin interactions.
\end{abstract}



\maketitle

\section{Introduction}

A single vacancy is the simplest intrinsic defect in a crystal, and has been seen in graphene with atomic resolution through, e.g. transmission electron microscopy (TEM)\cite{hash+04nature,meye+07nature} and scanning tunneling microscopy (STM).~\cite{uged+10prl} When the atom is removed, two scenarios are possible: either the disrupted bonds remain as dangling bonds or the structure undergoes a bond reconstruction through a Jahn-Teller rearrangement, and e.g. in 3D semiconductors we find a localized state and deep gap levels. Graphene on the other hand has notable 2D properties with the covalent bonding introducing two intrinsically different state types,  $\sigma$ and $\pi$, these last relevant for the Fermi-energy and Dirac point properties. The $\pi$-states are diffuse in the 2D planar ($x,y$) directions, but very localized on the $z$-direction with an in-plane node. As such, long range 2D electron-electron interaction is enhanced. In addition, the hexagonal structure with two sublattices creates for the $\pi$ states the special band structure with the Dirac point. We might thus expect special properties also for the vacancy in graphene. There is controversy from the experimental side about the reconstruction,~\cite{robe+13acsnano,zhan+16prl} however  a clear symmetry is found for the defect, and in particular from scanning tunneling microscopy~\cite{uged+10prl,zhan+16prl} it is found also that the defect level is resonant at the Dirac point, and induces magnetism.~\cite{zhan+16prl}

A number of theoretical studies of the electronic and magnetic properties of the vacancy in graphene have been reported in the past decade.~\cite{pere+06prl,mira+16prb,elba+03prb,leht+04prl,ma+04njp,yazy&helm07prb,pala&yndu12prb,casa+13prb, lin15jpcc,mene&capa15jpcm,rodr+16carbon,padm&nand16prb,zhan+16prl} In particular, first-principles calculations based on density functional theory (DFT) ~\cite{elba+03prb,leht+04prl,ma+04njp,yazy&helm07prb,pala&yndu12prb,casa+13prb, lin15jpcc,mene&capa15jpcm,rodr+16carbon,padm&nand16prb,zhan+16prl} yielded  widely varying results for the magnetic moment, in the range of $1.04-2.0$ $\mu_{B}$. For instance, Palacios and Yndur\'ain \cite{pala&yndu12prb} found that the magnetization decreases with decreasing defect density, tending to 1.0 $\mu_{B}$ in the low-density limit, in contrast to results reported by Yazyev and Helm\cite{yazy&helm07prb} where the magnetization increases from $1.15$ to $1.5$ $\mu_{B}$ with decreasing density, results that highlight the possibility of magnetic moment dependence with defect-defect interaction.

Regarding this last point, two typical approaches can be used for the simulation: model clusters, which are assumed to resemble the defect environment in the bulk, or periodic boundary conditions based on the choice of \textit{supercells} (SC). In the cluster model we must be careful about defect interaction with cluster edge states, which in the case of graphene can be critical.~\cite{fern&pala07prl,wang+08nl} As for the SC modeling, we must remember that we will study defects periodically arranged,~\cite{garc10prb} that is, we study an \textit{array of defects} that may induce spurious defect interactions.

Concerning the defect-edge interactions and focusing on the $\pi$-states, when we have flakes with zig-zag edges we can (depending on the flake symmetry) bring in Lieb's imbalance states~\cite{lieb89prl} that will group at the Fermi energy. These states are not realistic concerning the modeling of infinite graphene (no Lieb's imbalance), so we should not adopt such flakes for use as clusters. In the case of SCs we have for graphene three different symmetrical ($N\times N$) families, as shown in Refs. \onlinecite{aike+15prb,ding+11prb}, namely ($3n\times3n$), ($3n-1\times3n-1$) and ($3n+1\times3n+1$), where $n$ is an integer number. For the  $3n$ family, there occurs a folding of the $K-K'$ points onto the $\Gamma$-point of the SC Brillouin zone, that is, we will have degenerate, fully delocalized $\pi$-character states of different original symmetry crossing the Fermi energy at the SC $\Gamma$-point. These delocalized states interfere with localized defect states, through the long-range interaction property of the $\Gamma$-point, and is avoided when we adopt either one of the other families. Still for supercells, due to the $\pi$-symmetry of the relevant states at the Fermi region, we also have to take into account the possibility of long-range interaction between defects coming from parity in the zig-zag direction, as will be seen here.

In this work we adopt both the cluster approach, choosing hexagonal clusters with arm-chair and zig-zag edges, and periodic conditions with symmetrical cells from the different families ($3n\times3n$)($6\times6$), ($3n+1\times3n+1$)($7\times7$) and ($3n-1\times3n-1$)($8\times8$), as also a different symmetry cell ($6\times9$). We use semi-local DFT~\cite{perd+96prl} and hybrid DFT including a fraction $\alpha$ of Hartree-Fock exchange XC~\cite{perd+96jcp}, in which $\alpha$ is chosen to reproduce the properties of perfect graphene in the Fermi energy region.~\cite{pinh+15prb}

We find that, for the isolated vacancy defect, we can predict it introduces a magnetic moment of $2\mu_B$. It is critical to include a proper fraction of XC to arrive at a coherent description. Moreover, we find that periodic arrays of the defect can bring in interesting long-range spin dispersion effects.~\cite{just+14prb,ulyb-kats15prl}

\section{Methodology}\label{secmethodology}

All calculations in this work are performed using the all-electron FHI-aims code, \cite{blum+09cpc} with or without spin-polarization: the code employs numeric atomic orbitals obtained from \textit{ab-initio} all-electron calculations for isolated atoms, and can be used at the mean-field level with finite or infinite periodic models. The use of an all-electron code allows us to align the level structure of different simulation models by the deep 1$s^{2}$ Carbon orbitals. For dispersion interactions we adopt the Tkatchenko-Scheffler\cite{tkat&sche09prl} model which is sensitive to the chemical bonding environment. We employ \textit{tight} integration grids and \textit{tier2} basis sets,~\cite{havu+09jcp} and the atom positions are relaxed until the Hellmann-Feynman forces are smaller than 10$^{-3}$ eV/\AA{}. For periodic cells, we use the Monkhorst-Pack \cite{monk+76prb} ($\Gamma$-point included) scheme for sampling the Brillouin zone, with a [$6\times6\times1$] grid. The gaussian smearing is 0.01 eV for all calculations.

We compute the formation energy of the defect, and electronic and magnetic properties. The geometry optimization calculations are done using the PBE functional. The formation energy of a vacancy $E^{V}_{F}$ is calculated as

\begin{equation}
E^{V}_{F}=E(C_{n-1} H_{m})+E(carbon)-E(C_{n} H_{m})\,,\label{formation_energy}
\end{equation}
where here $E(C_{n} H_{m})$ is the total energy of the perfect cluster, $E(carbon)$ is the average energy of a single carbon atom in graphene, and $E(C_{n-1} H_{m})$ the total energy for the relaxed defect cluster. A similar computation is used in the case of periodic conditions.

Standard DFT, with exchange-correlation functionals in the local or semi-local (generalized gradient) approximations, is known to suffer from self-interaction errors \cite{cohe+08science}(SIE) leading to excessive delocalization of electrons.~\cite{kumm&kron08rmp} Hybrid density functionals reduce the SIE by mixing in a fraction of Hartree-Fock (HF) exchange and can significantly improve the study of many electronic properties. Now, most previous DFT calculations for the vacancy in graphene have been performed in the generalized gradient approximation. Since removing a carbon atom modifies the delocalized $\pi$-states into a localized defect state, a relevant improvement to the calculations could be the use of hybrid-DFT, since hybrid functionals may describe localized states better than GGA functionals.~\cite{pinh+15prb} The general form of the hybrid functional we use is

\begin{equation}
E_{x}^{PBEh} + E_{c}^{PBEh} = \alpha E_{x}^{HF}+(1-\alpha)E_{x}^{PBE} + E_{c}^{PBE}\,,\label{hPBE}
\end{equation}
where $E_{x}^{PBE}$ and $E_{c}^{PBE}$ denote the PBE exchange and correlation
energy, respectively, and $E_{x}^{HF}$ is the exact HF exchange energy: for example $\alpha$ = 0 corresponds to the PBE, and $\alpha$ = 0.25 to the PBE0 functional.~\cite{adam&baro99jcp}

The choice of the optimal $\alpha$ factor is system-dependent, as shown in Ref. \onlinecite{pinh+15prb}. There, the choice is based on Koopmans theorem for the ionization potential, gauged through many-body $G_0W_0$ calculations for finite systems. Here we simulate defects in graphene, and as such the goal is to find the optimal factor for the infinite extended system. We cannot apply the same procedure here since we need $GW$ results for the extended crystal, which cannot be obtained with the same code. We rely on literature theoretical results for the Fermi velocity $v_F$, coming from $GW$ methodology,\cite{trev+08prl,yang+09prl} and experimental results\cite{yu+09nl} for the Work Function $E_{W}$. With this rationale we choose $\alpha$ = 0.25, from the much-adopted functional PBE0, with which we obtain $v_F = 1.3\times 10^6 m/s$ and $E_{W}= 4.35 eV$ (compared to $v_F=0.98\times 10^6 m/s$ and $E_{W}= 4.24 eV$ with PBE).

Regarding the cluster simulation models, the $\sigma-\pi$ character allows us to use hydrogen-saturation of a graphene cut or nanoflake, with the required absence of imbalance states. We adopt for the electronic structure the same factor $\alpha$=0.25 since we are simulating by the cluster model the defect in the infinite crystal (it must be noted that the optimal factor~\cite{pinh+15prb} for this size of nanoflakes would be in the range $\alpha\sim 0.4-0.6$).

\subsection{Simulations Models}\label{clusterSC}

\begin{figure}
	\centering
	\includegraphics[width=0.8\textwidth]{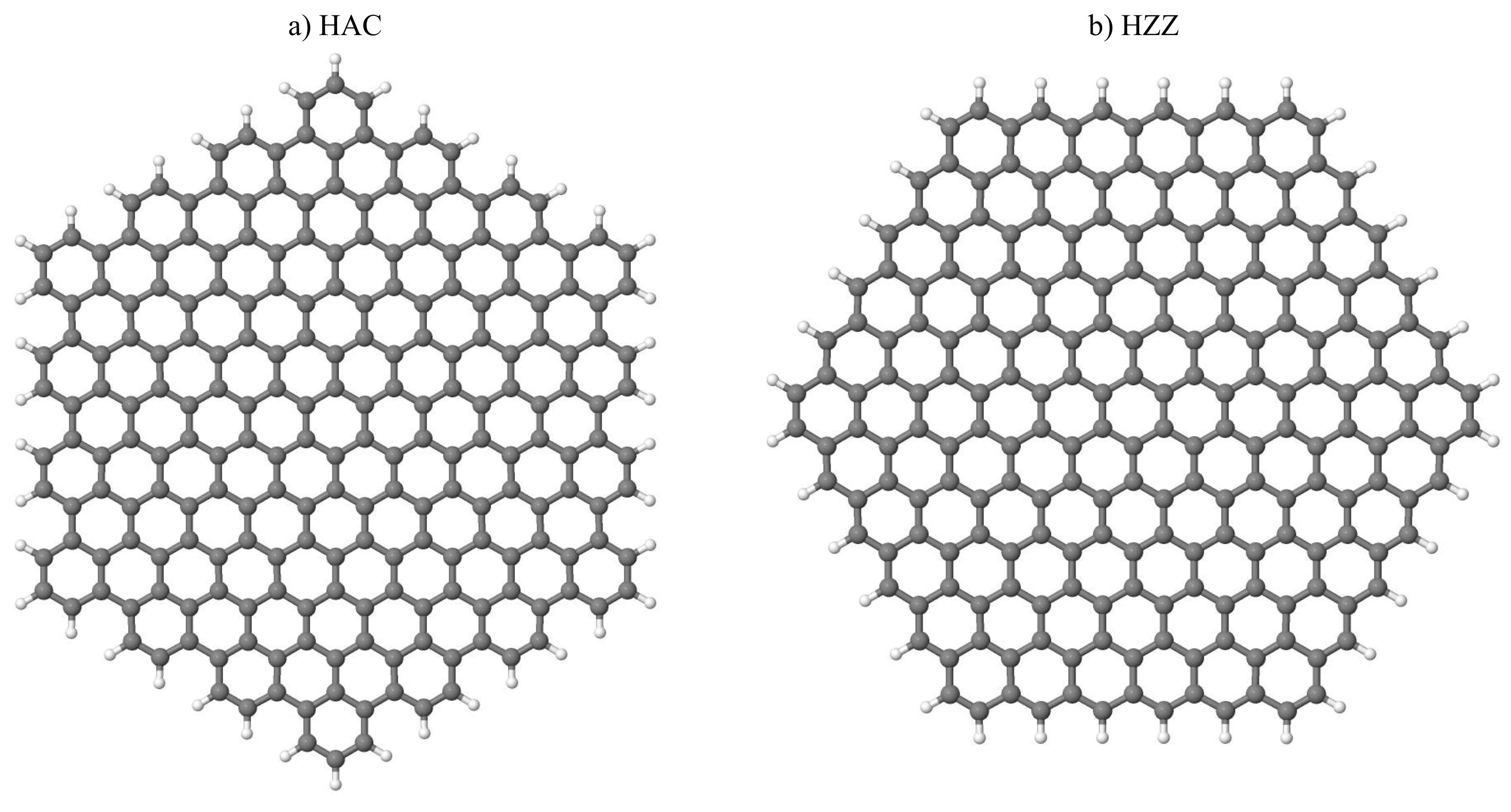}
	\caption{Cluster models adopted here for graphene, hexagonal $D_{6h}$ symmetry a) HAC $C_{222}H_{42}$ arm-chair and b) HZZ $C_{216}H_{36}$ zig-zag edges.}
	\label{GNFs_Perfect}
\end{figure}

It is possible to cut bulk graphene in different sizes and shapes, which allows us to create a cluster model which reproduces some relevant properties for the defect, such as symmetry. Graphene is a particular case for this approximation since nanoflakes can be specifically associated to chemically stable, well-known poly-aromatic hydrocabons PAHs, \cite{cocc+14jpca} and long flakes approach the well-studied graphene nanoribbons. \cite{jia+09science,cocc+12jpcl} In this last case, it is also well known that the character of the saturated border, arm-chair or zigzag, is very important for the electronic structure. \cite{enok12ps,fuji-enok13acr} We here will require that the structural conformation of the perfect cluster involves at least a $C_{3}$ symmetry operation, matching the $C_{3}$ rotation axis for regular graphene. Two series of hexagonal (H) clusters ($D_{6h}$ group) were analysed,  with arm-chair (AC) and zig-zag (ZZ) edges as shown in Fig. \ref{GNFs_Perfect} for (HAC)- $C_{222}H_{42}$ and (HZZ)-$C_{216}H_{36}$.

For the specific settings reported above, the resulting average carbon-carbon bond distances are 1.42 \AA{}, and 1.09 \AA{} for carbon-hydrogen bond lengths, and the (C-C-C) angle is $120^{\circ}$, thus we obtained the expected $sp^{2}$ hibridization character of carbon atoms. It is important to note that all structures were fully relaxed without any symmetry restriction. All structures remain completely flat after relaxation of atomic positions, regardless of the size, and the overall symmetry is maintained.

\begin{figure}
	\includegraphics[width=0.5\textwidth]{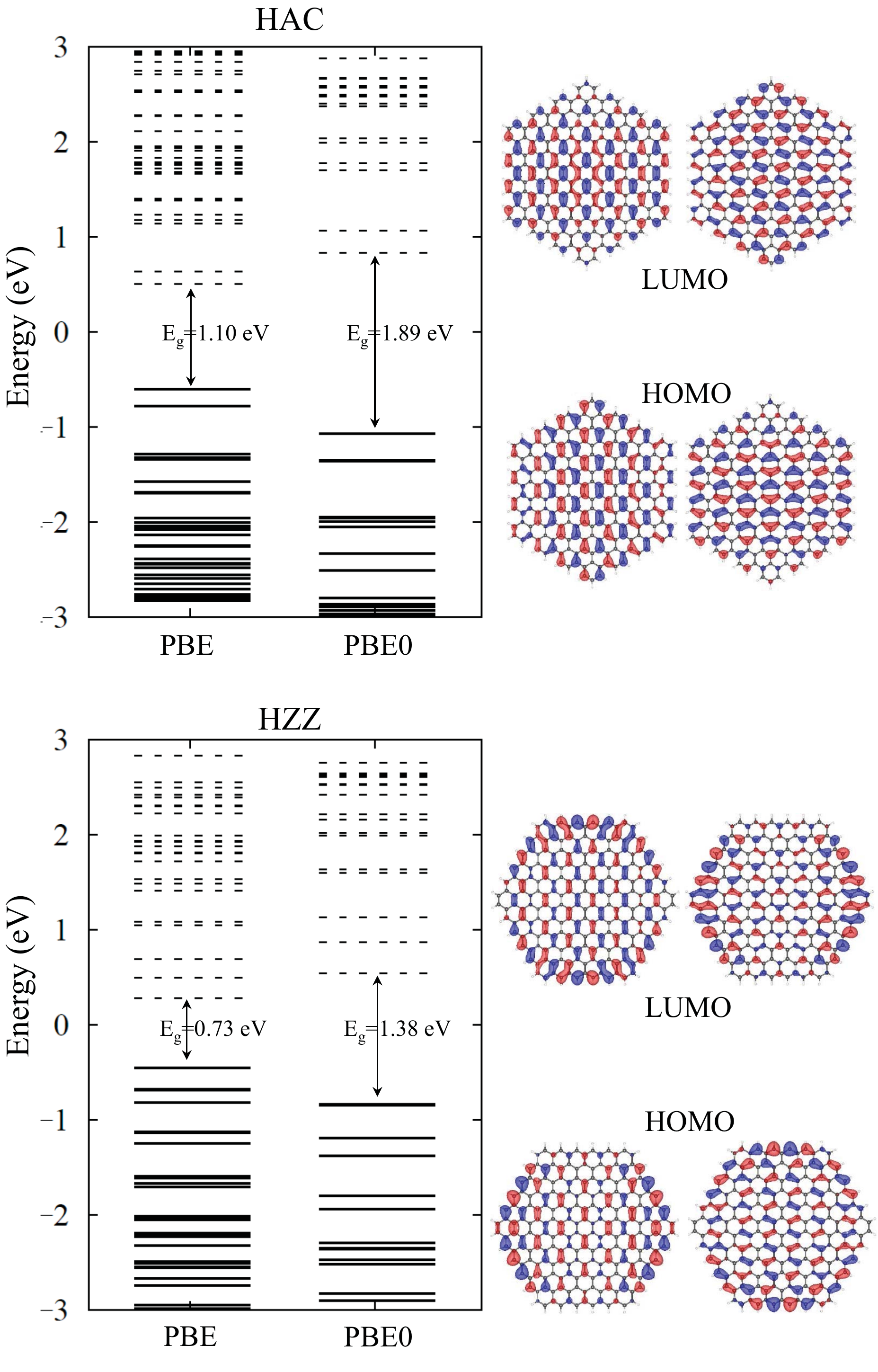}
	\caption{(Color online) Electronic energy levels for hexagonal H-clusters in the region near the Fermi level, results from PBE and PBE0. Solid (dotted) lines indicate occupied (unoccupied) states. Upper panel HAC-$C_{222}H_{42}$ and lower panel HZZ-$C_{216}H_{36}$. Energies aligned to the Fermi energy of the perfect crystal by the C-$1s^{2}$ average energy. Isosurfaces for the molecular orbitals (HOMO and LUMO) at the frontier energies from the PBE0 calculations.\label{KS_HGNFS}}
\end{figure}

To begin,  we show in Fig. \ref{KS_HGNFS} the energy spectra for the two clusters, which  have similar number of atoms, N $\sim{~220}$, but different edges. The highest occupied molecular orbital (HOMO) and the lowest unoccupied molecular orbital (LUMO) are two-fold degenerate for both HZZ and HAC. Already at the PBE level we have a sizeable HOMO-LUMO gap coming from the confinement effect, however it is worth noting that a quite significant increase is seen when we adopt PBE0.~\footnote{For the largest HAC cluster, and for the ($6\times9$) and ($8\times8$) supercells, using PBE0 we adopt \textit{tier2} with exclusion only of H(3d) basis; A H(3d) basis is anyhow included in the \textit{tier1} portion of basis.} As is well known \cite{baro+11acr,pusc-luft15jesrp} this significant increase in the HOMO-LUMO gap is seen for finite systems when a fraction of exact exchange is included via the hybrid functional approach; here these HOMO-LUMO gaps do not represent the actual gaps expected for nanoflakes, due to our choice of $\alpha$-factor, but just the cluster confinement effect. We simulated smaller clusters (HAC from 114 Carbon atoms and HZZ from 96 Carbon atoms) and we see that the HOMO-LUMO gap closure is very slow with cluster radius, as expected. An important characteristic of the frontier states in the case of zigzag edges is the concentration at the edges, not seen for the armchair cluster; however we can see that at the center, where the vacancy will be simulated, this effect should not be important. The difference in aromaticity between the clusters also bring a difference in the conjugation design of the frontier states, however both are fully conjugated.

As reported in the literature in two independent and almost simultaneous papers\cite{fern&pala07prl,wang+08nl} graphene nanostructures can have non-zero spin magnetic moment due to the sublattice imbalance mentioned above (Lieb's theorem).~\cite{lieb89prl} In the case of hexagonal clusters the sublattice imbalance does not exist, so the spin should be zero regardless of the type of edges. Indeed, we verify that none of the states show spin splitting, for both PBE and PBE0 functionals. Our results herein are in agreement with earlier theoretical results. \cite{fern&pala07prl,wang+08nl,wang+09prl}

Regarding the supercell (SC) models, as said above we will simulate the vacancy in the symmetrical $3n$ ($6\times6$) SC, much adopted in the relevant literature, however we will compare results with the $3n+1$ ($7\times7$)  and $3n-1$ ($8\times8$) SCs, which do not suffer from symmetry problems, and also for the non-symmetric ($6\times9$) SC.

\section{Vacancy in Graphene}\label{results2}

\begin{figure}
	\centering
	\includegraphics[width=0.8\textwidth]{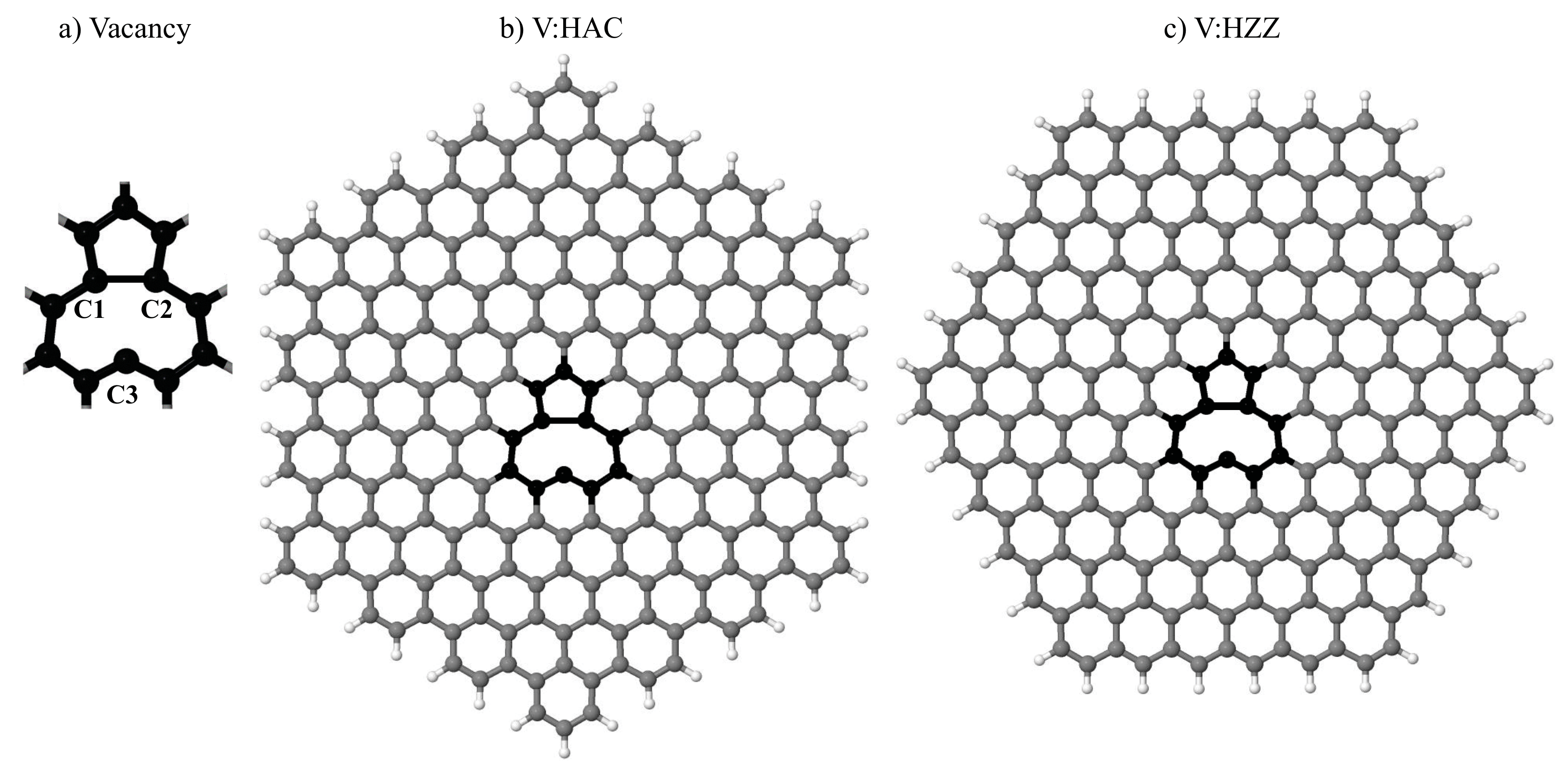}
	\caption{Vacancy in graphene a) visualization of the atoms in the close vicinity. Model clusters: fully optimized structures with spin-polarized PBE  b) VHAC-$C_{222}H_{42}$, c) VHZZ-$C_{216}H_{36}$. All structures remain flat after the structural relaxation.
		\label{V_GNFs}}
\end{figure}

Perfect graphene is non-magnetic, but the presence of the vacancy can induce magnetism, by breaking the symmetry of the $\pi$-electron system. Theoretical results are however not identical, and depend on the specific model or methodology adopted. As already commented in the Introduction, in the past years several works were dedicated to this study, using the different approaches of cluster or periodic conditions, and different theoretical formalisms.
\cite{pala&yndu12prb,elba+03prb,padm&nand16prb,ma+04njp,leht+04prl,mene&capa15jpcm,sing&krol09jpcm,wang&pant12prb,yazy&helm07prb,casa+13prb,zhan+16prl,rodr+16carbon,lin15jpcc,pere+06prl,mira+16prb}
Discussing first the results for the geometrical structure, there is consensus about the occurrence of Jahn-Teller distortion \cite{jahn&tell37prsla} for the surrounding atoms, with two of the three (here named C1 and C2, see Fig. \ref{V_GNFs}) reconstructing, and realizing a (weak) complete $\sigma$-$\pi$ bond, while the remaining (C3) atom carries the $\sigma$ and $\pi$ dangling bonds. It is also found that when the calculation is performed without spin-polarization, the C3 atom is projected out-plane, but when the spin is included the defect is back to full planar morphology. \cite{ma+04njp,leht+04prl,padm&nand16prb,casa+13prb}

Recently, it was proposed by Padmanabhan and Nanda \cite{padm&nand16prb} that indeed the two configurations are very close in energy, the out-of-plane (zero-spin) configuration being metastable, but of possible existence depending of the surrounding medium. This discussion leads us to the reported results for the magnetic moment of the defect, as obtained \cite{ma+04njp,yazy&helm07prb,padm&nand16prb,pala&yndu12prb,leht+04prl,casa+13prb,sing&krol09jpcm,mene&capa15jpcm,wang&pant12prb} from different works: Ma \textit{et al.} \cite{ma+04njp} and Lehtinen \textit{et al.} \cite{leht+04prl} reported a magnetic moment of $\sim{~1.04}$ $\mu_{B}$ ($8\times8$ super-cell, $3n-1$ family, PBE-sp); Palacios and Yndur\'ain \cite{pala&yndu12prb} detailed the analysis, and relate the magnetic moment of the vacancy with the size of used super-cells ($3n$ family). They found, using PBE-sp, already magnetic moment of $\sim {~1.7}$ $\mu_{B}$ in a $6\times6$ super-cell, but this value \textit{decreases} to $\sim {~1.0}$ $\mu_{B}$ as the supercell size increases ($15\times15$). In an earlier work (all three families, $4\times4$ to $12\times12$ super-cells, PBE-sp), Yazyev and Helm \cite{yazy&helm07prb} calculated however an \textit{increase} in the magnetic moment, from 1.12 to 1.53 $\mu_{B}$, when the distance between vacancies increases. In all cases reported with details, and as we will see below, the non-integer value of magnetic moment comes from the crossing of delocalized bending bands at the Fermi energy, giving the system a ``doped" character (which however is generated by a vacancy, a defect usually associated with a deep-level character in semiconductors). At the same time, these delocalized bands tell us that the interaction between defects in our supercell models may bring misleading effects.

\begin{figure*}
	\centering
	\includegraphics[width=1.0\textwidth]{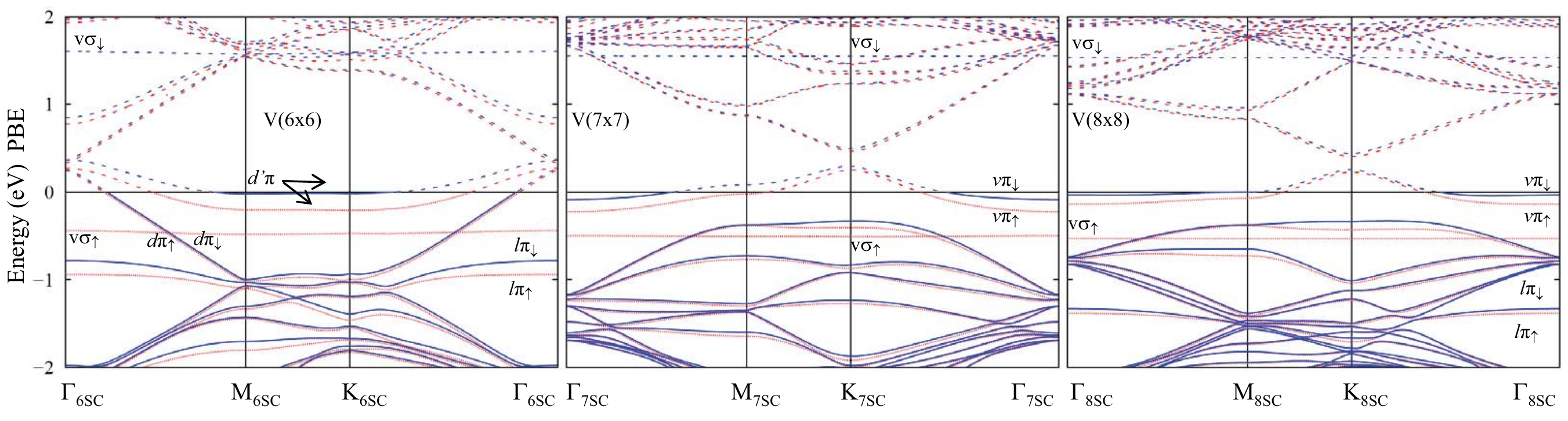}
	\caption{(Color online)Band structure for the vacancy defect in the region near the Fermi energy, results from spin-polarized PBE, for the supercells ($6\times6$) at left, ($7\times7$) at center and ($8\times8$) at right. Solid (dotted) lines indicate occupied (unoccupied) states. Energies aligned to the Fermi energy of the perfect crystal by the C-$1s^{2}$ average energy.}
	\label{v_level_band_PBE_sp}
\end{figure*}

At introducing a vacancy, in supercell or cluster models, we find that it exhibits a \textit{planar} Jahn-Teller distortion, and the point-group symmetry becomes $C_{2v}$. The reconstruction seen in Refs. \onlinecite{ma+04njp,leht+04prl,wang&pant12prb,padm&nand16prb,mene&capa15jpcm,pala&yndu12prb,yazy&helm07prb,sing&krol09jpcm} is recovered here, with two of the three affected carbon atoms binding to each other, and one single carbon with dangling bonds remains. The formation energy from PBE-sp results is approximately 7.6 eV in agreement with earlier theoretical results.  \cite{elba+03prb,kras+06cpl,sant+13cpl,skow+15csr,lin15jpcc}

\begin{figure}
	\centering
	\includegraphics[width=0.5\textwidth]{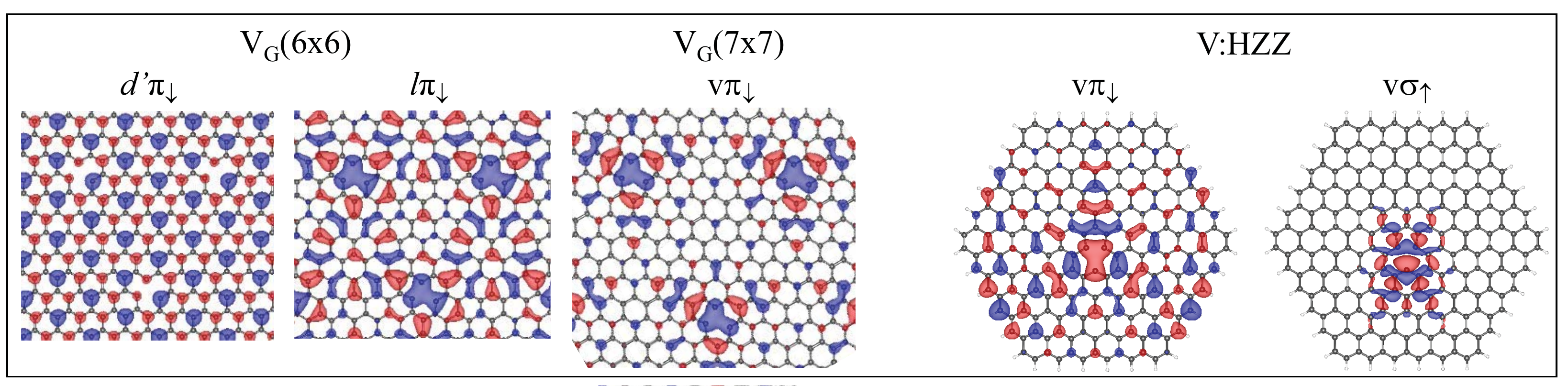}
	\caption{(Color online) Isosurfaces for the \textit{d'}$\pi$, \textit{l}$\pi$ and \textit{V}$\pi$  defect states (indicated  in Figs. \ref{v_level_band_PBE_sp} and  \ref{V_HAC_HZZ_spectra}) obtained with the spin-polarized PBE functional; the \textit{V}$\sigma$ state shown here for a cluster presents very similar character in all simulations, and the \textit{l}$\pi$ state is similar in the ($8\times8$) SC. }
	\label{isosurfs}
\end{figure}

We show next our results\footnote{From non-spin PBE results we observe that after optimization of atomic positions, the vacancy is reconstructed with a weak C1-C2 covalent bond (2.1 \AA{}) and the C3 is displaced out of the plane by $\sim {~0.5}$ \AA{}. Also we see that the surrounding atoms are displaced below the plane by $0.2$ \AA{}, resulting in a local rippling, similar to previously reported theoretical results. \cite{ma+04njp,kras+06cpl,skow+15csr,banh+11acsnano,rodr+16carbon} Moving to results from PBE \textit{with spin polarization}, after the relaxation defective graphene is completely flat. Here the spin-polarized flat structure is more stable by 0.07 eV} for the symmetric supercell models, for which the band structures are shown in Fig. \ref{v_level_band_PBE_sp}. We see first that in all cases we also find non-integer magnetic moments $\mu_{V}=1.49\mu_{B}$ for the ($6\times6$), $\mu_{V}=1.30\mu_{B}$ for the ($7\times7$) and $\mu_{V}=1.38\mu_{B}$ for the ($8\times8$) supercell, coming from the crossing of the bands at the Fermi energy. We stress however that the picture is qualitatively different when moving from the $3n$ to the $3n\pm1$ supercells. In the first case, the bands crossing the Fermi energy and seen in previous works\cite{pala&yndu12prb,casa+13prb,mene&capa15jpcm,rodr+16carbon,padm&nand16prb} are rather delocalized, not strictly defect-localized states. Indeed, while for the disruption of the $\sigma$-states we see quite localized defect states (flat bands) which we will call \textit{V}$\sigma$, with a sizeable spin-splitting ($\sim {~2}$ eV) in all models, the effect on the $\pi$-electrons for the ($6\times6$) SC is more spread-out and affects a numbers of states (or bands), in particular the folded bands from the (K, K') unit-cell points, that we call here \textit{d}$\pi$ and \textit{d'}$\pi$ (see Fig. \ref{isosurfs}). The vacancy-localized \textit{l}$\pi$ states are in this case affected first by the symmetry-folding and further by the parity of the supercell, interacting through the zig-zag connection; their influence on the final spin is not direct (both up- and down-spin states fully occupied) however the impact on the spin-density is seen. Looking now at our results for the ($7\times7$) SC, free from the symmetry-folding problems and parity connection, we see that delocalized bands are not involved anymore and the defect-related band which we will call \textit{V}$\pi$  is the one causing the final (non-integer) magnetic moment; as for the ($8\times8$) SC, we also have no symmetry-folding thus the \textit{V}$\pi$  state is the one crossing the Fermi energy, however we have parity connection and the \textit{l}$\pi$ states also contribute to the final magnetic moment.  Moving to the non-symmetrical ($6\times9$) SC, we still have symmetry folding to the $\Gamma$-point and the results present the same character found for the ($6\times6$), and a magnetic moment of $\mu_{V}=1.40\mu_{B}$. In summary, the defect-related states in these supercells show not only different total magnetic moment, but also very different character, and one cannot correlate the variations of $\mu_{V}$ to simple defect-defect distance, since these symmetry-related effects are very relevant.

\begin{figure}
	\centering
	\includegraphics[width=0.8\textwidth]{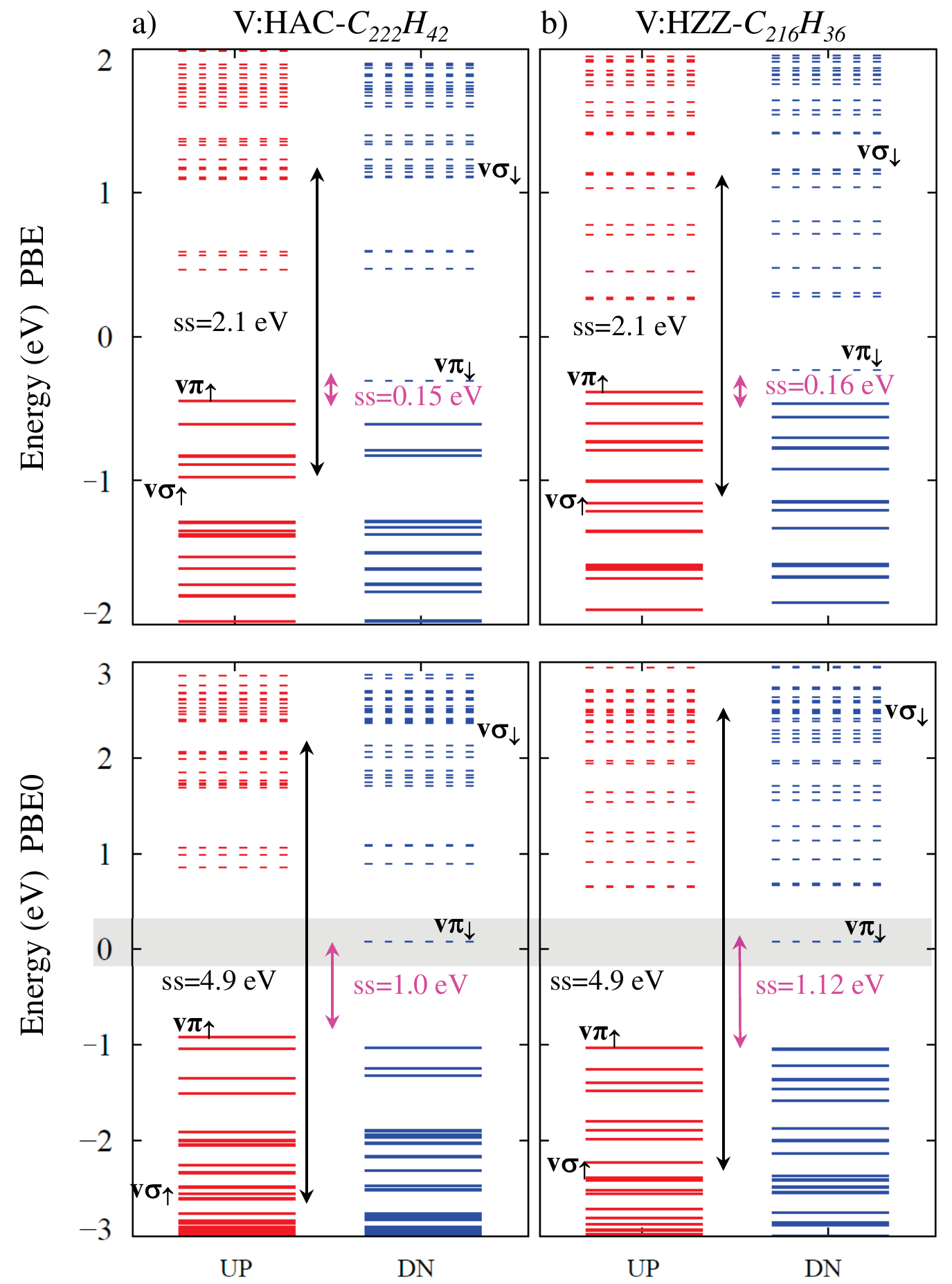}
	\caption{(Color online) Electronic energy levels for the vacancy in the clusters HAC-$C_{222}H_{42}$ (left) and HZZ-$C_{216}H_{36}$ (right), in the region near the Fermi energy. Results from spin-polarized PBE (top) and PBE0 (down) functionals. Solid (dotted) lines indicate occupied (unoccupied) states. Energies aligned to the Fermi energy of the perfect crystal by the C-$1s^{2}$ average energy.}
	\label{V_HAC_HZZ_spectra}
\end{figure}

We turn thus to our results using the cluster models. Fig. \ref{V_GNFs} shows the geometric structures of the vacancy in the  different clusters, optimized with PBE and spin polarization.  The cluster symmetry is broken, from $D_{6h}$ to $C_{2v}$. The removal of one $\pi$-orbital creates a lattice-imbalance in the hexagonal clusters, with direct effect on the magnetization. As was reported on Refs \onlinecite{wang+08nl,fern&pala07prl} perfect clusters with sublattice imbalance have non-zero spin magnetic moments, in accordance to Lieb's theorem. Considering this ``counting rules", the magnetic moment of the vacancy in graphene is predicted to be 1$\mu_{B}$. However, one has keep in mind that Lieb's theorem refers only to $\pi$ orbitals and the contribution from the $\sigma$-dangling orbital is not considered.

We here obtain a magnetic moment of 2$\mu_{B}$ already using PBE-sp in agreement with Wang and Pantelides. \cite{wang&pant12prb} This value of magnetic moment can be understood through the energy spectra in Fig. \ref{V_HAC_HZZ_spectra}: the defect states with different localized character are clearly identified, the lowest-energy occupied defect state \textit{V}$\sigma$ has higher localization, shows a spin-splitting of $\sim{~2.0}$ eV already at PBE level of theory, and contributes 1$\mu_{B}$ to the magnetization. We can see from Fig. \ref{isosurfs} that we recover here the defect state  \textit{V}$\pi$, very similar to that seen in the ($7\times7$) SC.  It is more spread over the cluster, the occupied spin orbital is the frontier HOMO level, and the spin-splitting is much lower $\sim{~0.2}$ eV than for the \textit{V}$\sigma$ states. Even so, the \textit{V}$\pi$ contribution is the same, and for the H-clusters the final magnetic moment is 2$\mu_{B}$.

At this point, we have conflicting results coming from the simulation of the same defect with the same formalism, just different theoretical models: from clusters we obtain $\mu_{V}=2\mu_{B}$ and for the SCs, as seen here and in the extensive literature mentioned above, PBE-sp results give a non-integer magnetic moment, where the spin splitting is complete for the localized $\sigma$ defect band, but a delocalized defect-induced $\pi$-band crosses the Fermi energy.

\begin{figure*}
	\centering
	\includegraphics[width=1.0\textwidth]{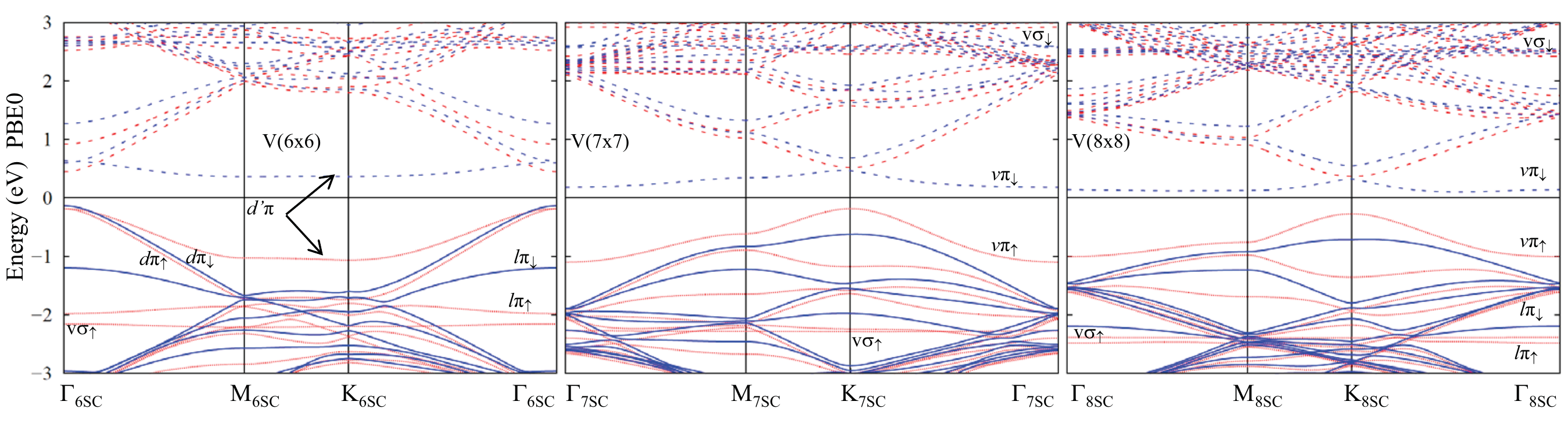}
	\caption{(Color online) Band structure for the vacancy defect in the region near the Fermi energy, results from spin-polarized PBE0 for the supercells ($6\times6$) at left, ($7\times7$) at center and ($8\times8$) at right. Solid (dotted) lines indicate occupied (unoccupied) states. Energies aligned to the Fermi energy of the perfect crystal by the C-$1s^{2}$ average energy.}\label{VSC_pbe0}
\end{figure*}

We now go to the final step of this work, which regards the effect of inclusion of Hartree-Fock exchange in the DFT functional.

In the case of hexagonal clusters the actual value of $\mu_{V}$ does not change $\mu_{V}=2\mu_{B}$, and we see in Fig.\ref{V_HAC_HZZ_spectra} that the main impact is the spin-splitting found for the defect levels, that for the \textit{V}$\pi$ state goes from $\sim{~0.2}$ eV to $\sim{~1.2}$ eV. The isosurfaces for these specific states in the HZZ cluster are shown in Fig. \ref{isosurfs} where we can see the distinct localization character of the $\sigma$ and $\pi$ states (similar characteristics are found for the HAC cluster). It is to be noted that using the PBE0 $\alpha$-fraction we observe, for all analysed clusters including the smaller ones, that the vacancy gives rise to a defect state ($V\pi$) indicated in Fig. \ref{V_HAC_HZZ_spectra} pinned at $E_{F}=0$, as seen by Ugeda et. al.~\cite{uged+10prl} and in accordance with previous theoretical predictions,~\cite{pere+06prl,pere+08prb} which is not the case using PBE. We pass next to the more impactant effect, seen for  all SCs and shown in Fig.\ref{VSC_pbe0}: we find that inclusion of the Hartree-Fock exchange eliminates the band-crossing at the Fermi energy in all supercells, enhancing the spin-splitting for the involved states and restoring the vacancy magnetic moment, $\mu_{V}=2\mu_{B}$.

Even if the magnetic moment is now the same, still for the ($6\times6$) SC, as also for the ($6\times9$), it comes from the splitting of the \textit{d'}$\pi$ levels, not from the defect-localized states, while in the case of the ($7\times7$) and ($8\times8$) SC the integer magnetic moment comes from the complete spin-splitting of the \textit{V}$\pi$ state. Indeed, from the ($7\times7$) to the ($8\times8$) SC both acceptor \textit{V}$\pi_\uparrow$ and donor \textit{V}$\pi_\downarrow$  levels approach the Fermi energy, each level showing a different localization character (band curvature close to the $K_{SC}$ point)  as detected in experimental results.~\cite{zhan+16prl}

\begin{figure}
	\centering
	\includegraphics[width=0.5\textwidth]{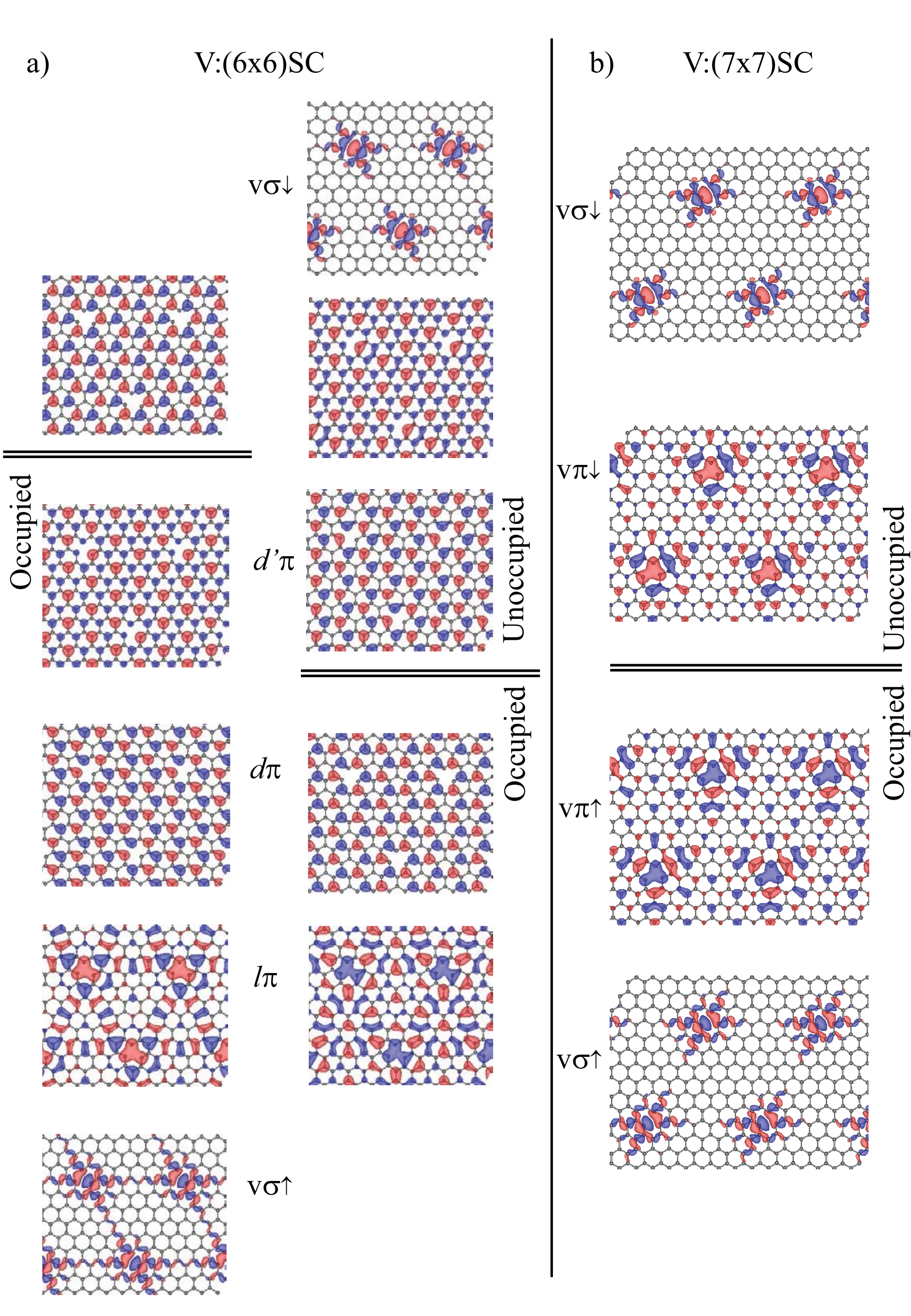}
	\caption{(Color online) Isosurfaces for relevant states of the vacancy in graphene obtained through PBE0 for the (a) $6\times6$ and (b) $7\times7$ supercells at the $\Gamma_{SC}$ point. In (a) spin-up states are shown at left and spin-down  at right; occupied and unoccupied levels are graphically separated by the double lines.}
	\label{isosurf6x6&7x7}
\end{figure}

Again the visualization of the isosurfaces for the relevant states in Fig. \ref{isosurf6x6&7x7} helps us in understanding this strong symmetry and exchange effect. Looking first at the impact of symmetry, we compare the isosurfaces for the defect states \textit{l}$\pi$ in the ($6\times6$) SC and the state \textit{V}$\pi$ in the ($7\times7$) SC; it can be seen that the ``triangular'' $C_{2v}$ character of the defect state seen for the cluster models (isolated defect) is reproduced in the later, with strong localization, while for the former we find defect-defect coupling coming from the parity connection imposed by the cell symmetry, on top of the band-folding of delocalized states that interact with the defect states.
All of this will define the effect of inclusion of exchange, since the localization character of each state will be relevant. While the \textit{V}$\sigma$ states have a strongly localized character and consequently strong spin-splitting, for the $\pi$-states the impact is much more subtle. In particular for the bands closer to the Fermi energy in the ($6\times6$) SC, due to the nodal character of these states the effect comes mostly in the energy stabilization, by the \textit{l}$\pi$  states, of the delocalized bands \textit{d}$\pi$. However, for the ($7\times7$) SC the impact of exchange is direct, as in the case of the cluster models, coming straight to the vacancy-related defect band.

\begin{figure}
		\centering
			 \includegraphics[width=0.5\textwidth]{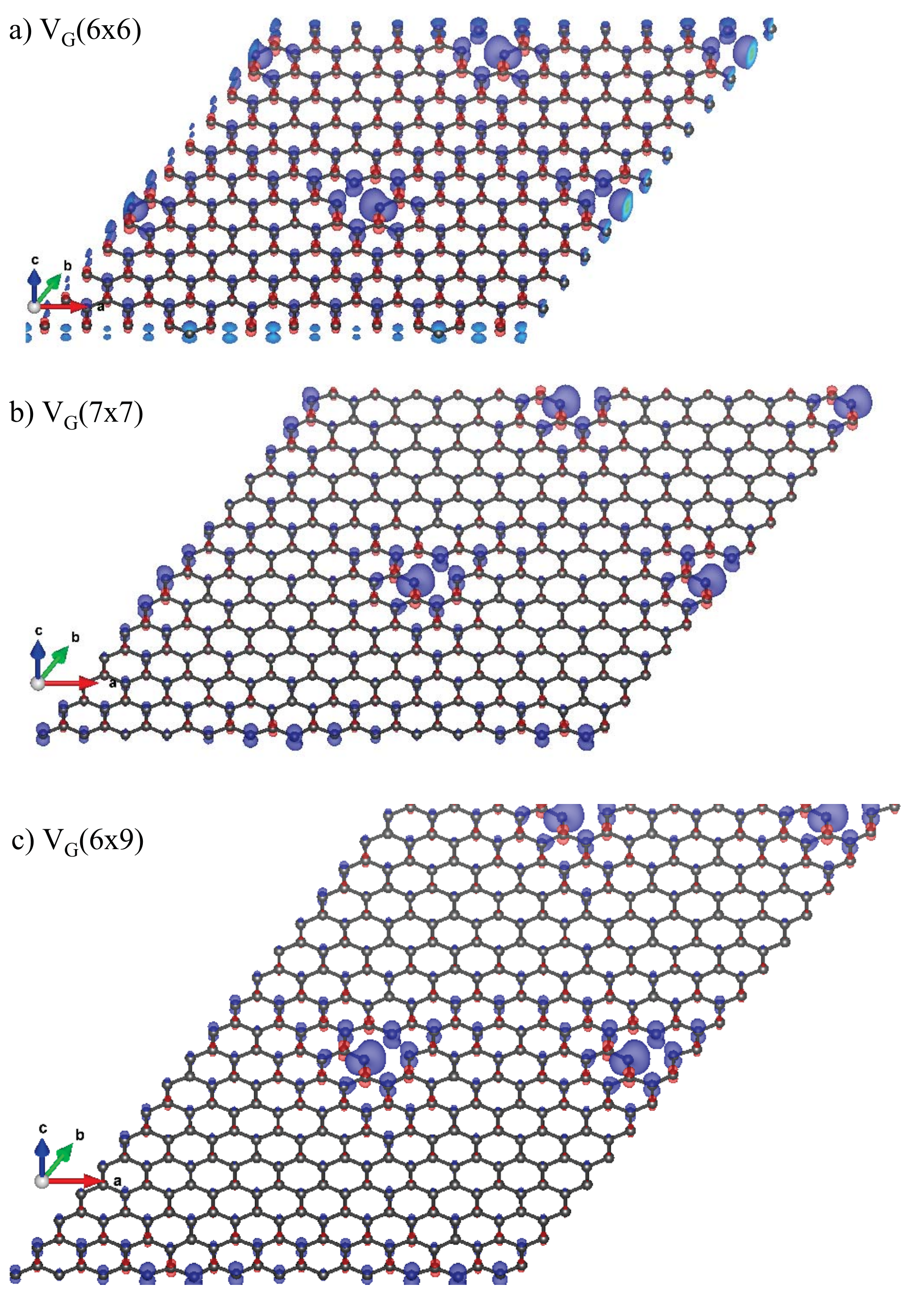}
		\caption{(Color online) Isosurfaces for the spin density ($0.05\AA{}^{-3}$) produced by the array of vacancies in graphene  obtained through PBE0 at the $\Gamma_{SC}$ point for (a) $6\times6$, (b) $7\times7$ and (c)  $6\times9$ SCs. }
		\label{vSC_PBE0_SD}
\end{figure}

Grouping our results from cluster and periodic boundary conditions, we see that with the inclusion of XC we can predict an integer magnetic moment of $\mu_{V}=2\mu_{B}$ for the isolated vacancy defect. The characteristic \textit{V}$\sigma$ level, seen in different DFT studies, shows a large spin splitting of very similar magnitude in our different simulations. For the defect $\pi$-states, we also see a characteristic acceptor level in the cluster and ($3n\pm1$) cells, pinned to the Fermi energy, responsible for the final integer magnetic moment. The confinement effect in the cluster models place the donor level much below, however from periodic conditions, in the ($3n\pm1$) cells, we see this level approaching the Fermi energy.

We turn now to the specific results obtained for the ($6\times6$) SC: the plot in Fig. \ref{vSC_PBE0_SD}, showing the \textit{spin density} across the cell, highlights the delocalized effect of this  $3n$-array of defects compared to the immediately one-unit larger ($7\times7$) SC. The high spin-density centered on the vacancy site comes from the difference in density between the \textit{l}$\pi$ up and down states, while the overall delocalization comes from the mixed $d'\leftrightarrow l$ character. We show also the spin density found for the ($6\times9$) SC, where we still see the same density along the zigzag direction in the shorter distance, while along the larger defect-defect distance, which has no parity connection, the density is much lower. We suggest this symmetry-derived behavior could be explored by designing chosen arrays of point defects.

\section{Summary and Conclusions}\label{secconclusions}

In summary, we have studied the vacancy defect in graphene through different approaches, and analysed the effects on the obtained electronic and magnetic structure. We used both cluster and periodic supercell models, and different exchange-correlation functionals, PBE and hybrid DFT-HF with a fraction of Hartree-Fock exchange $\alpha$ chosen to properly describe the electronic properties of graphene close to the Fermi energy, specifically PBE0 $\alpha=0.25$.
The results from the different simulation models show that, due to the specific symmetry and bilattice properties of graphene, symmetry-related coupling effects have to be carefully probed when using periodic boundary conditions to describe the isolated defect.

Our main conclusion is that that inclusion of the proper fraction of Hartree-Fock exchange is crucial for the description of the system, and allows us to arrive at the value of $\mu_{V}=2\mu_{B}$ for the magnetic moment of the isolated vacancy defect, after careful analysis of all adopted models. In addition, we find that the spin density created by an array of vacancies can show interesting directional properties.

\begin{acknowledgments}
This work was supported by FAPESP, INEO, and CAPES, Brazil and CONICYT, Chile. MJC acknowledges support from CNR-S3, Italy. We also acknowledge support on computer time  by the Res. Comput. Support Group (Rice University) and LCCA-USP, and NAP-NN-USP. We thank E. Molinari and L.G Dias da Silva for fruitful discussions.
\end{acknowledgments}

\bibliography{names,Vac_Graph_Biblio}

\end{document}